\documentstyle{article}

\newcommand{\beq}[1]{\begin{equation} \label{#1} }
\newcommand{\eeq}    {\end{equation}}

\newcommand{\oh}{\frac{1}{2}}
\newcommand{\th}{\frac{3}{2}}

\begin{document}

\title{T-noninvariant effect in muon capture\\
by $^6$Li nucleus with decay to a continuum}
\author{A.L.Barabanov\\
{\it The Kurchatov Institute, 123182 Moscow, Russia}}
\date{\mbox{}}
\maketitle

\begin{abstract}
The T-noninvariant transverse polarization of neutrons is
considered for muon capture by a $^6{\rm Li}$ nucleus with decay
into a quasistationary $2^+$ state of the three particle
$\alpha+{\rm n}+{\rm n}$ continuum. This polarization is
orthogonal to the plane spanned by the polarization axis of the
initial mesic atom and the neutron momentum. The situation in
which neutrons that are emitted in the plane orthogonal to the
axis of the mesic-atom polarization and which have oppositely
directed momenta equal in magnitude are selected is analyzed. The
wave function of the final state is constructed by using the
method of hyperspherical harmonics. In the approximation of the
allowed Gamow-Teller transition $1^+\to 2^+$, this neutron
polarization is expressed in terms of the T-noninvariant relative
phase of reduced matrix elements for transitions from the
$^6{\rm Li}$ ground state to various final-state configurations.
\end{abstract}

\section{Introduction}
\label{s1}

Evidences for violation of time reversal invariance (T-invariance)
are presently obtained only in K$^0$-meson decays. Attempts to
find T-noninvariant effects in nuclear reactions,
$\gamma$-transitions, $\beta$-decays of nuclei and free neutrons,
along with searches of electric dipole moments of nuclei and
elemental particles, have not been successful. However,
the results of
these investigations allowed to lower the upper limit of possible
mixture of T-noninvariant interactions to
$\sim 10^{-3}-10^{-4}$ (see, e.g.,~\cite{Herczeg_88}).

New approaches to T-invariance tests are of obvious interest. In
Refs.\cite{Deutsch_92}-\cite{Oziewicz_94} the possibilities of
T-invariance studies in modern experiments on muon capture were
considered. In particular, reactions with transition to continuum
$\mu+{\rm d}\to{\rm n}+{\rm n}+\nu_{\mu}$ and
$\mu+^3{\rm He}\to {\rm d}+{\rm n}+\nu_{\mu}$ were discussed in
Ref.\cite{Ciechanowicz_93}. It was pointed to the fact that
neutron polarization along $[{\bf n}_k\times {\bf n}_{\mu}]$,
where ${\bf n}_k$ and ${\bf n}_{\mu}$ are unit vectors along
neutron momentum and muon polarization, respectively, should be
sensitive both to T-noninvariant phases of formfactors in the week
semileptonic Hamiltonian and to T-invariance violation in
nucleon-nucleon interaction.

Of special interest are few body systems for study of
T-noninvariant components of nucleon-nucleon potentials. Only in
such systems reliable connection between experimental results and
constants of Hamiltonian may be established. Recently progress was
made towards microscopic description of $A=6$ nuclei as three
body systems $\alpha+{\rm N}+{\rm N}$ in the framework of
hyperspherical functions method (see, e.g.,
\cite{Danilin_89}-\cite{Danilin_93}). In particular,
in Ref.\cite{Shulgina_93} the calculation was performed of the
probability of muon capture by nucleus $^6{\rm Li}$ with
transition to the ground state of $^6{\rm He}$.

The binding energy of the nucleus $^6{\rm He}$ equals 0.975~MeV.
Above this value the three particle continuum
$\alpha+{\rm n}+{\rm n}$ begins. There is a resonance in this
continuum with spin and parity~$2^+$, energy 1.8~MeV and width
0.1~MeV. Transition into this resonance state as the result of
muon
capture by nucleus $^6{\rm Li}$ with spin and parity~$1^+$ is of
Gamow-Teller type. Thus the existence of the resonance simplify
the description of muon capture with decay to continuum. This
paper is devoted to estimation of T-noninvariant neutron
polarization along the direction
$[{\bf n}_k\times {\bf n}_{\mu}]$ after muon capture by $^6$Li
with decay to continuum.

\section{Muon capture --- general formalism}
\label{s2}

Non-relativistic transition Hamiltonian in nucleon space results
from relativistic Hamiltonian by Foldy-Wouthuysen transformation
(see, e.g., \cite{Eisenberg_70}) and is expressed as a power
series in the ratio $E_{\nu}/Mc^2$, where $E_{\nu}$ is neutrino
energy, and $M$ is nucleon mass. To estimate the effect
in a Gamow-Teller transition we restrict our consideration to
zero-order terms. Thus a dimensionless Hamiltonian is of the
form
\beq{2.1}
\hat{h}'=\sum_{j=1}^A
\exp{\left(-i\frac{{\bf p}_{\nu}{\bf r}_j}{\hbar}\right)}
\left(-ig_V{\rm b}^+_4(\sigma_{\mu},\sigma_{\nu})+
g_A{\bf b}^+(\sigma_{\mu},\sigma_{\nu})\hat{\vec{\sigma}}_j\right)
\hat{\tau}_-(j).
\eeq
We use here 4-vectors
\beq{2.2}
{\rm b}_{\lambda}(\sigma_{\mu},\sigma_{\nu})=
i\;\overline{u_{\mu}(0,\sigma_{\mu})}
\gamma_{\lambda}(1+\gamma_5)
u_{\nu}({\bf p}_{\nu},\sigma_{\nu}),
\eeq
which are introduced in accordance with Ref.\cite{Eisenberg_70}
and expressed in terms of muon and electron 4-spinors.
The matrixes $\gamma_{\lambda}$ are taken in pseudoeuclidian
metric. The Hamiltonian $\hat{h}'$ is a matrix in a space of spin
projections
of muon $\sigma_{\mu}$ and neutrino $\sigma_{\nu}$. It contains
formfactors of vector $g_V$ and axial-vector $g_A$ interactions,
depending on squared transferred 4-momentum
$k^2=k_{\lambda}k_{\lambda}$. Summation over all $A$ nucleons
enters in the Hamiltonian. Spin operator $\hat{\vec{\sigma}}_j$
and lowering isospin operator $\hat{\tau}_-(j)$ act in a space
of $j$-th nucleon. We have
\beq{2.3}
\hat{\tau}_-|p>=|n>, \qquad \hat{\tau}_-|n>=0.
\eeq
Nucleon coordinates ${\bf r}_j$ are reckoned from the
center-of-mass of the nucleus. In the non-relativistic
approximation a muon 4-spinor is of the form
\beq{2.4}
u_{\mu}(0,\sigma_{\mu})=
\left(\varphi_{\mu}(\sigma_{\mu}) \atop 0\right),
\eeq
where $\varphi_{\mu}(\sigma_{\mu})$ is a usual two-component
spinor.

The nucleus captures a muon from the $1s$-state. If nucleus spin
$J_i$ differs from zero, the state is splitted into two hyperfine
sublevels with angular momenta $F^{\pm}=J_i\pm 1/2$ and energies
$E^{\pm}$. Interference terms $\sim \exp{(\pm i(E^+-E^-)t/\hbar)}$
of differential capture probability oscillate rapidly and go to
zero on averaging over life time of mesoatom. So angular
correlations in muon capture should be calculated separately for
each hyperfine sublevel, even though these sublevels are not
distinguished in the experiment.

Let the wave function $\Psi_{J_iM_i}$ describes initial nucleus
in a state with projection $M_i$ of spin $J_i$ on an axis $z$. The
mesoatom state from which the muon is captured is determined
by the function
\beq{2.5}
|F>=\sum_{\xi}a_{\xi}(F)
\sum_{M_i\sigma_{\mu}}C^{F\xi}_{J_iM_i\oh\sigma_{\mu}}
\Psi_{J_iM_i}\varphi_{\mu}(\sigma_{\mu}).
\eeq
The amplitudes $a_{\xi}(F)$ contain an information on mesoatom
polarization. They are normalized to the unity
$\sum_{\xi}|a_{\xi}(F)|^2=1$, where $\xi$ is a projection of the
angular momentum $F$ on the polarization axis (axis $z$). Let us
introduce a unit vector ${\bf n}_{\mu}$ along this axis. Mesoatom
polarization is given as usual by the quantity
\beq{2.6}
p_1(F)=\frac{<\xi>}{F}, \qquad
<\xi>=\sum_{\xi}\xi|a_{\xi}(F)|^2.
\eeq

Let the final state of the system $\alpha+{\rm n}+{\rm n}$,
formed
after muon capture by the nucleus $^6{\rm Li}$, is described by
the function $\Psi^{\sigma_1\sigma_2}_f$, where $\sigma_1$ and
$\sigma_2$ are spin projections of 1-st and
2-nd neutrons on an axis
$z'$. Thus we have for the amplitude of probability to find
the 1-st neutron with spin projection $\sigma_1$ and
the 2-nd neutron with spin projection $\sigma_2$ on an axis $z'$
\beq{2.7}
a^F(\sigma_1,\sigma_2)\sim
\sum_{\xi}a_{\xi}(F)
\sum_{M_i\sigma_{\mu}}C^{F\xi}_{J_iM_i\oh\sigma_{\mu}}
<\Psi^{\sigma_1\sigma_2}_f|\hat{h}'|\Psi_{J_iM_i}>.
\eeq

The probability of finding of the 2-nd neutron summed over
non-observed spin projections of the 1-st neutron and neutrino is
proportional to the equation
\beq{2.8}
w^F_0\sim\sum_{\sigma_2}
\sum_{\sigma_1\sigma_{\nu}}|a^F(\sigma_1,\sigma_2)|^2.
\eeq
While for the averaged spin projection of the 2-nd neutron we
obtain
\beq{2.9}
w^F_1\sim\sum_{\sigma_2}\sigma_2
\sum_{\sigma_1\sigma_{\nu}}|a^F(\sigma_1,\sigma_2)|^2.
\eeq
Neutron polarization along an axis $z'$ is determined by the
ratio
\beq{2.10}
p^F_1=\frac{w^F_1}{w^F_0}.
\eeq
This polarization depends on angular momentum of mesoatom.

\section{Wave functions}
\label{s3}

At low excitation energies nuclei with mass number $A=6$ behave as
systems built up from three bodies --- $\alpha$-particle and two
nucleons. It is convenient to construct the wave functions in the
form of series on hyperspherical harmonics (see, e.g.,
\cite{Jibuti_93}). In particular in the framework of this method
the wave function was calculated of the $^6$Li ground state $1^+$
with isospin $T=0$ in Ref.\cite{Danilin_91}.

In Ref.\cite{Danilin_93} the same method was used to study the
structure of $2^+$ continuum state of the system
$\alpha+{\rm n}+{\rm n}$ with isospin $T=1$. Generally a
three-body continuum state is described in a center-of-mass
system by the function $\Psi_{{\bf p}_x{\bf p}_y}$, depending on
asymptotic momenta ${\bf p}_x$ and
${\bf p}_y$, which are conjugated
to normalized Jacobi coordinates ${\bf x}$ and ${\bf y}$.
In the system $\alpha+{\rm n}+{\rm n}$ these coordinates
\beq{3.1}
{\bf x}=\sqrt{\frac{M}{2}}({\bf r}_{n2}-{\bf r}_{n1}), \qquad
{\bf y}=\sqrt{\frac{4M}{3}}\left({\bf r}_{\alpha}-
\frac{{\bf r}_{n1}+{\bf r}_{n2}}{2}\right),
\eeq
are proportional to the relative radius-vectors of two neutrons
$\vec{\rho}_{12}={\bf r}_{n2}-{\bf r}_{n1}$ and of the
$\alpha$-particle with respect to the center-of-mass of the
nucleon pair
$\vec{\rho}_3={\bf r}_{\alpha}-({\bf r}_{n1}+{\bf r}_{n2})/2$.
The momenta ${\bf p}_x$ and
${\bf p}_y$ are expressible in
terms of neutron and $\alpha$-particle momenta in the laboratory
system
\beq{3.2}
{\bf p}_x=\frac{1}{\sqrt{2M}}({\bf p}_{n2}-{\bf p}_{n1}), \qquad
{\bf p}_y=\frac{1}{\sqrt{3M}}\left(\frac{{\bf p}_{\alpha}}{2}-
({\bf p}_{n1}+{\bf p}_{n2})\right).
\eeq

Hyperspherical harmonics $\Phi^{l_xl_y}_{KLM}(\Omega)$ depend on
five variables
$\Omega=(\theta,\theta_x,\varphi_x,\theta_y,\varphi_y)$ and are
fixed by five quantum numbers --- hypermoment $K$, orbital momenta
$l_x$, $l_y$, total orbital momentum $L$ and its projection $M$ on
an axis $z$. Hyperspherical harmonics are of the form
\begin{eqnarray}
\Phi^{l_xl_y}_{KLm}&=&N^{l_xl_y}_K
(\sin{\theta})^{l_x}(\cos{\theta})^{l_y}
P^{l_x+\oh l_y+\oh}_n(\cos{2\theta})\times{}
\nonumber\\[\medskipamount]
&&{}\times\sum_{m_xm_y}C^{Lm}_{l_xm_xl_ym_y}
Y_{l_xm_x}(\theta_x,\varphi_x)
Y_{l_ym_y}(\theta_y,\varphi_y),
\label{3.3}
\end{eqnarray}
where $P^{\alpha\beta}_n$ are Jacobi polynomials, variable $n$
takes the values $n=0,1,2\ldots$, hypermoment $K$ equals
$2n+l_x+l_y$, and $Y_{lm}$ are usual spherical harmonics.
Normalized factor is determined by the formula
\beq{3.4}
N^{l_xl_y}_K=\left(
\frac{2n!(2n+l_x+l_y+2)\Gamma(n+l_x+l_y+2)}
{\Gamma(n+l_x+\th)\Gamma(n+l_y+\th)}\right)^{1/2}.
\eeq

Let us introduce wave vectors
${\bf k}_x={\bf p}_x/\hbar=(k_x,\theta_{kx},\varphi_{kx})$,
${\bf k}_y={\bf p}_y/\hbar=(k_y,\theta_{ky},\varphi_{ky})$,
where polar $\theta_{kx}$ and azimuth $\varphi_{kx}$ angles
specify the direction of the vector ${\bf k}_x$, as well as the
angles $\theta_{ky}$ and $\varphi_{ky}$ give the direction of
${\bf k}_y$. Variables $k$ and $\theta_k$ are determined as follow
\beq{3.5}
k=\sqrt{k^2_x+k^2_y},\qquad
k_x=k\sin{\theta_k},\qquad
k_y=k\cos{\theta_k}.
\eeq
Note, that the energy in the center-of-mass system equals
\beq{3.5.2}
E=\hbar^2k^2/2.
\eeq
So parameter $\theta_k$ determines how this energy is distributed
between the subsystems ${\rm n}+{\rm n}$ and $\alpha+(2{\rm n})$.
In according with Ref.\cite{Danilin_93} the continuum wave
function of $\alpha+{\rm n}+{\rm n}$ may be written as series on
the states with definite angular momenta $J_f$ and their
projections $M_f$ on an axis $z$
\beq{3.5.3}
\Psi^{\sigma_1\sigma_2}_{{\bf p}_x{\bf p}_y}=
\sum_{J_fM_f}\sum_{K\gamma}\sum_{m\sigma}
C^{J_fM_f}_{LmS\sigma}C^{S\sigma}_{\oh\sigma_1\oh\sigma_2}
\left(\Phi^{l_xl_y}_{KLm}(\Omega_k)\right)^*
\Psi^{K\gamma}_{J_fM_f}.
\eeq
Here $S$ is the total spin of two neutrons, and the index $\gamma$
denotes a set $\{l_xl_yLS\}$. The expansion coefficients contain
the hyperharmonics, depending on variables
$\Omega_k=(\theta_k,\theta_{kx},\varphi_{kx},\theta_{ky},
\varphi_{ky})$.

Calculations of Ref.\cite{Danilin_93} have shown that the
continuum resonance state $2^+$ of the $\alpha+{\rm n}+{\rm n}$
system is formed mainly by two terms of the above expansion,
corresponding to the same hypermoment $K=2$. The contribution of
the component $\gamma1=\{0220\}$ in the internal region varies
from 45 to 70 \% depending on the calculation method, while the
contribution of the component $\gamma2=\{1111\}$ changes from
35 to \mbox{20 \%}. We
shall restrict our consideration to these two terms in the
expansion (\ref{3.5.3}) for the wave function of the final state.

The wave functions $\Psi_{J_iM_i}$ and
$\Psi^{\sigma_1\sigma_2}_{{\bf p}_x{\bf p}_y}$ are, of course,
antisymmetrized over transposition of space, spin and isospin
coordinates of two nucleons of the system
$\alpha+{\rm N}+{\rm N}$. Therefore matrix element of Hamiltonian
(\ref{2.1}) $\hat{h}'=\hat{h}'_1+\hat{h}'_2$ is equal to the sum
of two identical matrix elements of the operators $\hat{h}'_1$ and
$\hat{h}'_2$. In this paper we take into account only the
contribution of the allowed s-wave matrix element to the amplitude
of Gamow-Teller transition $1^+\to 2^+$. Introducing the reduced
matrix elements in accordance with Ref.\cite{Balashov_78}, we get
for the matrix element of spherical component of the operator
$\hat{\vec{\sigma}}$
\beq{3.6}
<\Psi^{K\gamma}_{J_fM_f}|
\hat{\sigma}_{\alpha}|\Psi_{J_iM_i}>\simeq
\frac{4\pi}{\sqrt{3}}C^{J_fM_f}_{J_iM_i1\alpha}[101]_{K\gamma}.
\eeq
Spherical components are determined by usual rule
$\hat{\sigma}_{\pm 1}=
\mp (\hat{\sigma}_x\pm \hat{\sigma}_y)/\sqrt{2})$,
$\hat{\sigma}_0=\hat{\sigma}_z$.

If T-invariance holds, thus the phases of the wave functions may
be chosen in accordance with the standard condition
\cite{Bohr_69}
\beq{3.7}
\hat{T}|JM>=(-1)^{J+M}|J-M>,
\eeq
where $\hat{T}$ is the time reversal operator. In this case the
reduced matrix elements $[101]_{K\gamma}$ are real. In another way
the coefficients of expansion of the wave functions
$\Psi_{J_iM_i}$ and $\Psi^{K\gamma}_{J_fM_f}$ on hyperspherical
harmonics are hyperradial functions which satisfy the set of
coupled differential equations of second order. The reduced matrix
elements are the one-dimension integrals of these functions.
Following the rule (\ref{3.7}), one may make real all hyperradial
functions and, consequently, reduced matrix elements.

With nucleon-nucleon potentials violating time reversal
invariance the set of
coupled differential equations have no real solutions. The rule
(\ref{3.7}) does not hold. So the reduced matrix elements become
complex. The procedure of calculating of imaginary T-noninvariant
corrections to functions satisfying the set of coupled
differential equations of second order was considered in
Ref.\cite{Barabanov_95}. In this paper we express the neutron
polarization induced by time reversal violation in terms of
imaginary parts of matrix elements.

\section{Analysis of T-noninvariant effect}
\label{s4}

In the framework of our formalism the neutron polarization
(\ref{2.10}) depends on three momenta --- on neutrino momentum
${\bf p}_{\nu}$, which is equal by magnitude and opposite by
direction to the momentum of the center-of-mass of the system
$\alpha+{\rm n}+{\rm n}$, and on Jacobi momenta ${\bf p}_x$ and
${\bf p}_y$. In principle, these three momenta may be determined
by measuring the momenta ${\bf p}_{n1}$, ${\bf p}_{n2}$ and
${\bf p}_{\alpha}$ in the laboratory coordinate system. However,
such experiment is non-realistic.

As the other limiting case one may consider the situation when the
momentum of only one neutron is measured, while the integration
over all possible momenta of the other neutron, $\alpha$-particle
and neutrino holds. In such experiment it is impossible to control
the decay into the resonance state $2^+$. Therefore this case is
not interesting for us. Note in addition, that recalculation of
the effect, expressed in terms of natural for three-body problem
Jacobi momenta ${\bf p}_x$, ${\bf p}_y$, to laboratory momenta
${\bf p}_{n1}$, ${\bf p}_{n2}$  and ${\bf p}_{\alpha}$ needs the
cumbersome numerical procedure. Such recalculation would
complicate the interpretation of observable effect.

The optimal experiment should be not-too-compex on setting, on
the one hand, and simple on interpretation, on the other hand. We
consider the following situation. Lets assume that the momenta of
two neutrons are detected in the experiment. We select the events
when these momenta are equal by magnitude and opposite by
direction. Thus the magnitude and direction of Jacobi momentum
${\bf p}_x$ is determined by equation (\ref{3.2}). Furthermore,
the total momentum of two neutrons equals zero, therefore in
accordance with formula (\ref{3.2}) the momentum ${\bf p}_y$
coincides by direction with the $\alpha$-particle momentum.
Besides, due to the fact that the total momentum of four particles
in the final state equals zero
\beq{4.1}
{\bf p}_{n1}+{\bf p}_{n2}+{\bf p}_{\alpha}+{\bf p}_{\nu}=0,
\eeq
in the selecting cases ${\bf p}_{\alpha}=-{\bf p}_{\nu}$. This
means that the direction of Jacobi momentum ${\bf p}_y$ is exactly
opposite to the neutrino momentum. Thus the integration over all
non-observable direction ${\bf p}_y$ and ${\bf p}_{\nu}$ easily
performs. To control the decay into the resonance $2^+$ continuum
state one need to measure only the total energy of
the $\alpha$-particle. In addition the integration over the
parameter $\theta_k$ holds if the distribution of the total energy
between the subsystems ${\rm n}+{\rm n}$ and $\alpha+(2{\rm n})$
is not taken into account.

We normalize the continuum wave functions of the system
$\alpha+{\rm n}+{\rm n}$ as in Ref.\cite{Danilin_93}. Thus we
obtain the following expression for the differential probability
(\ref{2.8}) of detecting of two neutrons with opposite momenta
(\mbox{${\bf p}_{n1}=-{\bf p}_{n2}$}) and with Jacobi momentum
${\bf p}_x$ in the solid angle $d\Omega_x$ for the Gamow-Teller
transition $J_i\to J_f=2$
\beq{4.2}
dw^F_0=A_{\mu}|g_A|^2\frac{2J_f+1}{2J_i+1}C(J_i,F)
\left(|[101]_{\gamma1}|^2+|[101]_{\gamma2}|^2\right)
\frac{d\Omega_x}{4\pi}\frac{\Delta\Omega_y}{4\pi}dk.
\eeq
Here we use the constant
\beq{4.3}
A_{\mu}=\lambda_{\mu}
\frac{8R(Z)Z^3}{3}
\frac{(E_{\nu}/m_{\mu}c^2)^2}
{(1+E_{\nu}/E_f)(1+m_{\mu}c^2/E_i)^3},
\eeq
where
\beq{4.4}
\lambda_{\mu}=\left(\frac{e^2}{\hbar c}\right)^3
\frac{(G\cos{\theta_C})^2(m_{\mu}c^2)^5}
{{\strut \hbar}^7 {\strut c}^6}\simeq
1.005 \cdot 10^3 \quad {\mbox s}^{-1},
\eeq
$G$ is the week constant, $\theta_C$ is the Cabbibo angle,
$m_{\mu}$ is the muon mass, $Z$ is the charge of the initial
nucleus, and $R(Z)$ is the correction for nucleus size
\cite{Balashov_78}. Neutrino energy $E_{\nu}$ is determined by the
total energy $E_f$ of the final system $\alpha+{\rm n}+{\rm n}$,
which includes rest masses of the particles, and by the total
energy $Q_{\mu}$, released in muon capture, as follow
\beq{4.5}
E_{\nu}=E_f\left[\left(1+
\frac{2Q_{\mu}}{E_f}\right)^{1/2}-1\right]\simeq
Q_{\mu}\left(1-\frac{Q_{\mu}}{2E_f}+\ldots\right).
\eeq
Differential $dk$ is related with differential $dE$ of the final
system energy according the formula (\ref{3.5.2}):
$dk=dE/(\hbar\sqrt{2E})$, where $E$ and $dE$ should be taken equal
to the energy and width of the resonance state $2^+$. The solid
angle $\Delta\Omega_y$ fixes the value of the possible deviation
of the momentum ${\bf p}_y$ from the direction, opposite to
neutrino momentum. The coefficient $C(J_i,F)$ depends on spin
$J_i$ of initial nucleus and on total angular momentum of mesoatom
$F=J_i\pm 1/2$ and is of the form
\beq{4.6}
C(J_i,F)=1+\sqrt{6}U(F\oh J_i1,J_i\oh)U(2J_i11,1J_i),
\eeq
where $U(abcd,ef)=((2e+1)(2f+1))^{1/2}W(abcd,ef)$ is the
normalized Racah function \cite{Jahn_51}.

Note, that the differential probability (\ref{4.2}) does not
contain the terms like $\sim \cos \theta$, where $\theta$ is the
angle between the momentum ${\bf p}_x$ and the mesoatom
polarization axis ${\bf n}_{\mu}$. This is associated with
the identity of
emitting neutrons and, consequently, with ambiguity of the choice
of the direction ${\bf p}_x$. There exists also the other special
feature of the coefficient $C(J_i=1,F)$. Let $J_i=1$, thus
$F=1/2$ or $3/2$. It is easy to check that $C(J_i=1,F)$ equals
zero, if the total angular momentum of the initial state is
$F=1/2$. This is the result of angular momentum conservation.
Indeed, the expression (\ref{4.2}) is obtained in the
approximation of the s-wave neutrino emission with respect to the
center-of-mass of the system $\alpha+{\rm n}+{\rm n}$. It is clear
that as neutrino spin equals $1/2$ and the spin of the final
resonance state is $J_f=2$, the total angular momentum in the
final state is 3/2 or 5/2. So this state can not be obtained from
the initial state with $F=1/2$.

We turn now to the situation when neutron momenta
${\bf p}_{n1}=-{\bf p}_{n2}$ are perpendicular to the polarization
axis ${\bf n}_{\mu}$ of the initial mesoatom. By convention we
assign a number 1 to one neutron and a number 2 to the other
neutron and fix the direction of the vector ${\bf p}_x$. According
to the equation (\ref{3.2}) this vector coincides by direction
with momentum ${\bf p}_{n2}$. Choosing an axis $z'$ along
$[{\bf p}_x\times {\bf n}_{\mu}]$, we calculate the polarization
of the 2-nd neutron along the axis $z'$ using the formulas
(\ref{2.8})-(\ref{2.10}). We get
\beq{4.7}
p^F_1=\frac{32}{27\pi}\sqrt{\frac{6}{5}}p_1(F)
\frac{D(J_i,F)}{C(J_i,F)}
\frac{{\rm Im}\,\left([101]_{\gamma 1}[101]^*_{\gamma 2}\right)}
{|[101]_{\gamma 1}|^2+|[101]_{\gamma 2}|^2}.
\eeq
Under a different choice of neutron numbering we change only the
direction of the vector ${\bf p}_x$ and, consequently, of the axis
$z'$. Thus we should interpret the result (\ref{4.7}) as
polarization of any of two identical neutrons along
$[{\bf n}_k\times {\bf n}_{\mu}]$, where ${\bf n}_k$ is the unit
vector along the neutron momentum. The expression obtained
contains the spin factor
\begin{eqnarray}
&&D(J_i,F)=\left(\frac{(2J_i+1)F}{F+1}\right)^{1/2}
\sum_{E=0,2}(2E+1)U(121E,12)\times{}
\nonumber\\[\medskipamount]
&&\hspace{-1cm}{}\times\left[U(\oh J_iF1,FJ_i)\left\{
\begin{array}{ccc}
2&J_i&1\\
2&J_i&1\\
E&  1&1
\end{array}\right\}-
\left(\frac{2F+1}{5}\right)^{1/2}
\left\{\begin{array}{ccc}
J_i&F&1/2\\
J_i&F&1/2\\
E  &1&1
\end{array}\right\}U(1J_i2E,2J_i)+{}\right.
\nonumber\\[\medskipamount]
&&{}+2\left(\frac{(2F+1)(2J_i+1)}{3}\right)^{1/2}
\sum_{G=0,2}(2G+1)^{1/2}
\left\{\begin{array}{ccc}
J_i&F&1/2\\
J_i&F&1/2\\
G  &1&1
\end{array}\right\}\times{}
\nonumber\\[\medskipamount]
&&\hspace{3cm}\left.{}\times
\sum_{H=0,2}(2H+1)U(11HE,G1)
\left\{\begin{array}{ccc}
2&J_i&1\\
2&J_i&1\\
E&G  &H
\end{array}\right\}\right].
\label{4.8}
\end{eqnarray}
It equals zero when $J_i=1$, $F=1/2$ for the same reason as
$C(J_i,F)$. For $J_i=1$, $F=3/2$ we have $C(1,3/2)=1.5$,
$D(1,3/2)=0.237$.

We see that the neutron polarization (\ref{4.7})
differs from zero if
the phases of the reduced matrix elements $[101]_{\gamma1}$ and
$[101]_{\gamma2}$ do not coincide. Let us remind that if
T-invariance holds the phases of wave functions may be chosen
according the rule (\ref{3.7}). In this case all reduced matrix
elements are real. Denoting T-noninvariant phase of the product
$[101]_{\gamma 1}[101]^*_{\gamma 2}$ by $\phi$, where according
Ref.\cite{Herczeg_88} $\phi<10^{-3}$, we get a rough estimate of
the effect
\beq{4.10}
p^F_1\simeq 0.1\phi p_1(F)
\frac{\left|[101]_{\gamma2}/[101]_{\gamma1}\right|}
{1+\left|[101]_{\gamma2}/[101]_{\gamma1}\right|^2}.
\eeq
Note, that the polarization $p_1(F)$ of mesoatom $\mu+^6{\rm Li}$
in the state $F=3/2$ is about 0.05 \cite{Favart_70}.
However, in the experiments with previously polarized atoms (see,
e.g., \cite{Kuno_87,Newbury_91}) the quantity $p_1(F)$ may be
considerably higher.

\section{Conclusion}
\label{s5}

In this paper T-noninvariant transverse polarization is considered
for neutrons from reaction of muon capture by $^6{\rm Li}$ nucleus
with decay to resonance $2^+$ continuum state of three particles
$\alpha+{\rm n}+{\rm n}$. This polarization is normal to the plane
formed by polarization axis of initial mesoatom and neutron
momentum. We analyze the situation when selected neutrons have
equal and
opposite directed momenta which are perpendicular to the axis of
mesoatom polarization.

To estimate T-noninvariant effect we use an explicit expression
for the wave function of the final state in the hyperspherical
harmonics method. An analysis of the structure of $2^+$ resonance,
performed in Ref.\cite{Danilin_93}, allowed us to take into
account only two leading terms in this wave function. Only the
allowed matrix element of Gamow-Teller transition $1^+\to 2^+$ was
considered. In these approximations the neutron polarization is
expressed in terms of T-noninvariant relative phase of reduced
matrix elements for transitions from the ground state
of $^6{\rm Li}$
nucleus into the different configurations of the final state.

The current level of microscopic description of nuclei $A=6$ as
three-body systems $\alpha+{\rm N}+{\rm N}$ in the method of
hyperspherical harmonics \cite{Danilin_89}-\cite{Danilin_93} allow
to calculate the reduced matrix elements. We intend to perform
such calculations as the next step. This will enable to relate the
effect considered with parameters of nucleon-nucleon potentials
violating time reversal invariance (see, e.g., \cite{Gudkov_92}).
\bigskip

The author is grateful to Yu.V.Gaponov, B.V.Danilin and
N.B.Shul'gina for
useful discussions. This work was supported by International
Science Foundation, grant number is M7C300.

\end{document}